\documentclass{aa}
\usepackage{epsfig}
\usepackage{epic}
\def \eg {{\it e.g.} }
\def \etal {{et al.} }
\def \ie {{\it i.e. } }

\newcommand{\new}[1]{{{#1}}}

\begin{document}

\title{FARGO~: a fast eulerian transport algorithm for differentially
rotating disks}

\author{F.~Masset}

\offprints{F.~Masset \new{(F.S.Masset@qmw.ac.uk)}}

\institute{School 
of Mathematical Sciences, Queen Mary \& Westfield College, Mile End Road, London E1 4NS,
United Kingdom
}
\thesaurus{02.01.2, 02.08.1, 03.13.4}

\maketitle

\begin{abstract}

We present an efficient and simple modification of the standard
transport algorithm used in explicit eulerian fixed polar grid
codes, aimed at getting rid
of the average azimuthal velocity when applying the Courant condition.
This results in a much larger timestep then the usual procedure, and it
is particularly well-suited to the description of a Keplerian disk where
one is traditionally limited by the very demanding Courant condition 
on the fast orbital motion at the inner boundary. In this modified
algorithm, the timestep is limited by the perturbed velocity and by
the shear arising from the differential rotation. FARGO stands for
``Fast Advection in Rotating Gaseous Objects''. The speed-up resulting
from the use of the FARGO algorithm is problem dependent. In the
example presented here, which shows the evolution of a Jupiter sized
protoplanet embedded in a minimum mass protoplanetary nebula,
the FARGO algorithm is about an order of magnitude 
faster than a traditional transport scheme, with a much smaller numerical
diffusivity.

\end{abstract}

\section{Introduction}

We want hereafter to model the hydrodynamical (HD)
evolution of a disk described on a fixed polar eulerian grid. For the
sake of simplicity we are only going to deal with a two dimensional
Keplerian disk, but the algorithm can be extended with little additional
effort to any gaseous thin or thick disk in differential rotation.
Usually in this kind of numerical simulations the timestep is limited
by the Courant Friedrich Levy (CFL) condition at the inner boundary, 
where the motion is fast and the cells are narrow. Indeed, the 
ratio of the distance swept by the material in one timestep to the
cell width must be lower than unity over the whole grid, otherwise
a numerical instability occurs (i.e.
non physical short-wavelength oscillations appear, grow exponentially
and spoil the model). In a Keplerian disk this ratio (which we
call hereafter the CFL ratio) decreases as
$r^{-3/2}$. Since in most cases the ``interesting region'' of the
grid is located much further than the grid inner boundary, the CFL
ratio in the region of interest
is much smaller than unity, which corresponds to a waste of
computing time, and, as we are going to see below, to an enhanced
undesirable numerical viscosity.
The most obvious solution to get rid of such a limitation is to
work in the comoving frame. Unfortunately, most finite-difference
HD eulerian codes require an orthogonal system of coordinates
(Stone \& Norman, 1992), which makes them unsuitable if one wants
to work in the comoving frame in a differentially rotating disk,
and even a non-orthogonal grid eulerian code would be unable to track 
accurately
the fluid motion  after a few orbits, due to the strong winding
of the coordinate system.
On the other hand, one can adopt a Lagrangian description of the
disk (Whitehurst, 1995), but the implementation is much more tricky
and difficult. Furthermore, the geometry of an accretion disk provides
a polar mesh as a natural grid. We describe hereafter a simple
method which enables one to work on a fixed polar grid and to get
rid of the CFL condition on the average azimuthal velocity at each
radius.

\section{Notations and \new{standard} method}
\label{sec:clas}
We consider a polar grid composed of $N_s$ sectors, each one
$\Delta\theta=\frac{2\pi}{N_s}$ wide, and $N_r$ rings, with separations
at radii $R_{i_{(0\leq i\leq N_r)}}$. The inner boundary is then located
at the radius $R_0$, and the outer one at the radius $R_{N_r}$.
The density (and the internal
energy if needed by the equation of state) is centered in the
cells, and is denoted $(\Sigma_{ij})_{(i,j)\in[0,N_r-1]\times[0,N_s-1]}$.
The radial velocity is denoted $v^r_{ij}$, and is considered centered
in azimuth and half-centered in radius (applied at radius $R_i$, i.e.
at the interface between the cells $[i,j]$ and $[i-1,j]$). In a similar
way, the azimuthal velocity is denoted $v^\theta_{ij}$, and is
considered centered in radius and half-centered in azimuth
(i.e. at the interface between the cells $[i,j]$ and $[i,j-1]$~; 
throughout this paper 
the algebra on the $j$ coordinate is meant in $Z/N_sZ$
to account for the periodicity in azimuth).
Usually in a finite difference code the timestep is split in two main
parts (Stone \& Norman 1992). The first part 
is composed of  eulerian substeps which consist
in updating the HD quantities through the source terms
in the evolution equations, and which include all the physical
processes at work: pressure, gravity, viscosity, etc., and which
can formally be described by the transformation $\xi
\stackrel{E}{\rightarrow}\xi^a$, $\xi$ being any HD
field on the grid. The second part is the
transport substep, in which the quantities are conservatively
moved through the grid according to the flow $[(v_{ij}^r)^a, 
(v_{ij}^\theta)^a]$, and which can be formally represented
as $\xi^a\stackrel{R}{\rightarrow}\xi^b\stackrel{T}{\rightarrow}\xi^+$,
where $\xi^+$ denotes any HD field after a whole timestep is
completed, and $R$ and $T$ denote respectively the radial and
azimuthal transport operators, which can be alternated every other
timestep. The CFL condition comes both from the source part
and the transport part,
and the most stringent restriction is given by the $T$-substep,
due to the unperturbed azimuthal flow.
Classically, the azimuthal transport can be written as:
\begin{equation}
\label{eqn:clastrans}
\xi_{ij}^+=\xi_{ij}^b+\frac{\Delta t}{\Delta y_i}
(\xi_{ij}^{b,*/v^{\theta a}}v_{ij}^{\theta a}-
\xi_{ij+1}^{b,*/v^{\theta a}}v_{ij+1}^{\theta a})
\end{equation}
where $\Delta y_i=\frac{R_i+R_{i+1}}{2}\Delta\theta$ is the ``mean
azimuthal width'' of a cell. Eq.~(\ref{eqn:clastrans})
expresses the balance of the arbitrary
conservative quantity $\xi$ in the cell $[i,j]$
by computing the difference of
its inflow at the $[i,j-1]/[i,j]$ interface
with the velocity $v_{ij}^{\theta a}$ and its outflow
at the $[i,j+1]/[i,j]$ interface with the velocity $v_{ij+1}^{\theta a}$.
Actually we consider the flux of the upwinded interfacial quantity 
$\xi^{b,*/v^{\theta a}}$, where the ``$*/v^{\theta a}$'' operator depends
on the numerical method (donor cell, van Leer, PPA, see e.g. Stone
\& Norman 1992) and on the velocity field $v^{\theta a}$.

\section{New azimuthal transport algorithm}

\subsection{Overview}
\label{referee1}

\new{
Let us take as an example the angular momentum conservation equation:
\begin{equation}
  \label{eq:angmom1}
    \frac{\partial J}{\partial t}
+\underbrace{\frac{1}{r}\frac{\partial (v^\theta J)}{\partial\theta}}_{\mbox{azim. transport}}
+\underbrace{\frac{1}{r}\frac{\partial(rv^r J)}{\partial r}}_{\mbox{rad. transport}}
=\mbox{ Source terms }  
\end{equation}
where $J=\rho rv^\theta$. The transport equation of any
HD quantity $\xi$ will look the same as the
L.H.S. of eq.~(\ref{eq:angmom1}).

Now without loss of generality we can rewrite eq.~(\ref{eq:angmom1})
as:
\begin{equation}
  \label{eq:angmom2}
    \frac{\partial J}{\partial t}
+\frac{1}{r}\frac{\partial [(v^\theta-u) J]}{\partial\theta}
+\frac{u}{r}\frac{\partial J}{\partial \theta}
+\frac{1}{r}\frac{\partial(rv^r J)}{\partial r}
=\mbox{ Source terms }  
\end{equation}
where $u$ can be any quantity which does not depend on $\theta$. No
assumption has been made on the behavior of $J$ up to this point, and
eq.~(\ref{eq:angmom1}) and~(\ref{eq:angmom2}) are strictly equivalent.
If we take $u$ to be the average azimuthal velocity $\overline v^\theta$, 
then 
eq.~(\ref{eq:angmom2}) can be described as a composition of
different steps, and each of them can be worked out 
independently with  the well-known operator splitting technique:
\begin{itemize}
\item a source step,
\item a radial transport step,
\item an azimuthal transport step with the
velocity $v^\theta-\overline v^\theta$, which we are going to call the
azimuthal
{\em residual} velocity, 
\item and an additional step which corresponds
to the following PDE:

\begin{equation}
  \label{eq:shift1}
    \frac{\partial J}{\partial t}
+\frac{\overline v^\theta}{r}\frac{\partial J}{\partial \theta}
=0
\end{equation}
It is an easy matter to check that the solution of this last equation
can be written in a general way as:
\begin{equation}
  \label{eq:shiftsol}
  J(\theta,t)=J\left(\theta-\frac{\int_0^t\overline v^\theta dt}{r}
,0\right)
\end{equation}
which means that the solution of this equation at any time $t$ looks
like the initial profile ($t=0$), except for a shift 
$-\int_0^t\overline v^\theta dt/r$ in azimuth. It should be noted
that this is true {\em whatever the profile of $J$}, which can even contain
discontinuities (\ie shocks). 
In particular {\em no assumption has to be made} on the
linearity of the flow (\ie on the relative amplitude of the perturbed
quantities).

\end{itemize}

A qualitative reason of why such a decomposition is valid
 is that the time evolution
of the HD quantities can be described either by an observer sitting on a 
ring of radius $r$ which rotates at any instant in 
time with the average azimuthal 
velocity, or by an observer at rest in an inertial frame.
Now the time evolution of the system is of course observer-independent,
which is why their observations are reconciled through the simple shift
described by eq.~(\ref{eq:shiftsol}).

The idea on which the FARGO algorithm is based on is precisely to
evolve the HD quantities through operators which mimic in a discrete way
the different terms of eq.~(\ref{eq:angmom2}). The source step, the
radial transport step 
and the residual azimuthal
velocity transport step are performed in a standard
way (see \eg Stone \& Norman, 1992).
Now the last step in the operator-splitting described above, which
corresponds to a simple shift which amounts to be $\overline v^\theta
\Delta t/r$ in one timestep, can be implemented in such a way that 
the matter can sweep an arbitrary number of cell widths in one timestep.

In order to lay down the basic mechanism by which FARGO works, let us
take the following concrete 
example. We assume that, after the classical substeps
(which are the source step, the radial transport and the residual
azimuthal velocity transport), the material 
at a given radius $r$ has to be shifted
by $4.7$~cells in one timestep
(which means that $\overline v^\theta\Delta t/r\Delta\theta
=4.7$). What is actually done is that $4.7$ is 
decomposed as $4.7=-0.3+5$, \ie the nearest integer and a remainder
which by construction is lower or equal to $0.5$ in absolute value.
In the first substep of this shift step the material is shifted by this
remainder (here $-0.3$), 
which can be achieved through a classical transport method since
the remainder is lower or equal to $0.5$ in absolute value (it has to
be $\leq 1$ in order for the standard transport method to be possible),
with the additional simplicity that the corresponding velocity field
is uniform (which is actually why shift and transport happen to coincide
in this special case, since there is no compression in the 
corresponding flow).
The second substep just corresponds to an integer number
of cells shift, which is done in our example simply by copying the
content of cell $j$ into cell $j+5$, for any $j$.

A more formal and detailed description of the FARGO algorithm is 
given in the next section.

}

\subsection{Mathematical formulation of each step of the FARGO algorithm}

In the modified algorithm, the azimuthal transport substep is
split in several parts. We assume that the timestep $\Delta t$
has already be chosen, and defer discussion of the timestep
constraints until section~\ref{subsec:ts}.
We first compute the average azimuthal
velocity at each radius:
\begin{equation}
\label{new:step1}
\overline{v}_i^\theta=\frac{1}{N_s}\sum_{j=0}^{N_s-1}
v_{ij}^{\theta a}
\end{equation}
We then introduce the residual velocity: $v_{ij}^{\theta {\rm res}}
=v_{ij}^{\theta a}-\overline{v}_i^\theta$, and the ``shift number'' at
each radius~:
\begin{equation}
\label{new:step2}
n_i=E\left[\overline{v}^\theta_i\frac{\Delta t}{\Delta y_i}
\right]
\end{equation}
where $E[X]$ denotes the nearest integer to the real $X$.
We define the constant residual velocity to be:
\begin{equation}
\label{new:step3}
v_i^{\theta{\rm cr}}=\overline{v}^\theta_i-n_i\frac{\Delta y_i}{\Delta t}
\end{equation}
Hence the total velocity can be expressed as:
\begin{equation}
\label{new:step4}
v_{ij}^{\theta a}=v_i^{\theta {\rm SH}}+v_i^{\theta {\rm cr}}+
v_{ij}^{\theta {\rm res}}
\end{equation}
where the ``shift velocity'' $v_i^{\theta {\rm SH}}=
n_i\frac{\Delta y_i}{\Delta t}$ corresponds
to a uniform shift of $n_i$ cells over one timestep.

We first transport the HD quantities according to the flow
$v^{\theta {\rm res}}$~:
\begin{equation}
\label{new:step5}
\xi_{ij}^c=\xi_{ij}^b+\frac{\Delta t}{\Delta y_i}
(\xi_{ij}^{b,*/v^{\theta {\rm res}}}v_{ij}^{\theta {\rm res}}-
\xi_{ij+1}^{b,*/v^{\theta {\rm res}}}v_{ij+1}^{\theta {\rm res}})
\end{equation}
then to the uniform flow $v^{\theta {\rm cr}}$~:
\begin{equation}
\label{new:step6}
\xi_{ij}^d=\xi_{ij}^c+\frac{\Delta tv_i^{\theta {\rm cr}}}{\Delta y_i}
(\xi_{ij}^{c,*/v^{\theta {\rm cr}}}-
\xi_{ij+1}^{c,*/v^{\theta {\rm cr}}})
\end{equation}
We split \new{the first part of} 
the transport into two parts ($v^{\theta {\rm res}}$ and
$v^{\theta {\rm cr}}$) instead of using a single transport step with
the velocity $v^{\theta {\rm res}}+v^{\theta {\rm cr}}$, in order
to ensure (as can be checked below given the timestep constraints)
that in each of these transport substeps the material sweeps at most
half a cell \new{(it could sweep up to one cell, but for reasons which
will become clear in section~\ref{sec:1d}, we prefer to take a half cell
limitation), and in order for the continuity considerations of
section~\ref{sec:cont} to apply}.
Finally, the quantities are transported along the $v^{\theta {\rm SH}}$
uniform flow~:
\begin{equation}
\label{new:step7}
\xi_{ij}^+=
\xi^d_{ij-n_i}
\end{equation}
Only the first two parts of this transport step introduce some
numerical diffusion. The last one, \new{given by eq.~(\ref{new:step7})}, 
which in many
cases corresponds to the largest part of the motion, does not introduce
any numerical error, since it just corresponds to a circular permutation
of the grid cells, \new{or in other words it is just an integer discrete
version of the shift given by eq.~(\ref{eq:shiftsol}).}

\new{A precise quantification of the lower numerical diffusivity
of FARGO is beyond the scope of this paper though. An extremely
rough estimation can be done in the case of the comparison of
a standard method (in which the effective CFL ratio is a sizable
fraction of one) and a FARGO method for which $n_i\ne 0$. If we
assume that numerical effects will behave in azimuth as a physical
viscosity would do, then the effective numerical viscosity in
FARGO is about $n_i/C_0$ times lower than the standard method's one,
where $C_0$ is the CFL standard dimensionless limitation factor,
which is detailed in the next section. Nevertheless a variety of
numerical experiments can be found below which all show that
FARGO's numerical diffusivity is smaller than the standard method's.}

\subsection{Timestep limitation}
\label{subsec:ts}
In the \new{standard} transport method, the timestep limitation arises
from the combination of four different constraints (see e.g.
Stone \& Norman 1992), namely the
fact that a flow advected test particle in cell $[i,j]$ should
not sweep a distance longer than $\Delta y_i$
in azimuth nor longer than $R_{i+1}-R_i$ in radius over one
timestep (which introduces the limit timestep $\delta t_2$
and $\delta t_3$ in Stone \& Norman's paper), 
and that the wavefront of any wave present in the system
should not travel across a whole cell over one timestep
(Richtmyer \& Morton, 1957), which corresponds to the
limit timestep $\delta t_1$ in Stone \& Norman's paper. 
The last
constraint comes from a stability limit arising from the viscosity 
(numerical or physical). With the modified azimuthal transport
algorithm, the constraint on the azimuthal motion has to be 
modified slightly. Following Stone \& Norman's notation, instead
of writing $\delta t_3^{ij}=\Delta y_i/v^{\theta a}_{ij}$,
we write~:
\begin{equation}
\label{eqn:timelimit}
\delta t_{3}^{ij}
=\frac{\Delta y_i}
{v_{ij}^{\theta a}-\overline v_i^\theta}
=\frac{\Delta y_i}
{v_{ij}^{\theta \rm res}}
\end{equation}
which means that the timestep limitation  comes now from the perturbed
azimuthal velocity, 
which results in a much higher absolute 
value of $\delta t_3$.
Another limitation arises from the shear. Indeed we do not want the shear
to disconnect the two neighboring cells
$[i,j]$ and $[i+1,j]$ after one timestep. We write this condition as~:
\begin{equation}
\label{eqn:shearlimit}
\delta t_{\rm shear}^{ij}
={1 \over 2}{\left(
\frac{v_{ij}^{\theta a}}{\Delta y_i}
-
\frac{v_{i+1j}^{\theta a}}{\Delta y_{i+1}}
\right)^{-1}
}
\end{equation}
Following Stone \& Norman's notations, we finally adopt~:
\[\Delta t = C_0/\{\max_{ij}[(\delta t_1^{ij})^{-2}+
(\delta t_2^{ij})^{-2}+
(\delta t_{3}^{ij})^{-2}+
\]

\begin{equation}
\label{eqn:timestep}\phantom{\Delta t = C_0/\{\max_{ij}[}
(\delta t_4^{ij})^{-2}+
(\delta t_{\rm shear}^{ij})^{-2}]^{1/2}\}
\end{equation}

\subsection{Continuity}
\label{sec:cont}

At each timestep, $N_r$ values of $n_i$ (with $i\in[0,Nr-1]$), used in 
eq.~(\ref{new:step7}),
are computed using eq.~(\ref{new:step2}). These integer values scale
roughly as $R_i^{-3/2}$. The shift on the central parts generally
amounts to several cells over one timestep, while in the outer
parts $n_i$ is small, and possibly zero. One can wonder whether
or not problems may arise at the radii $R_i$ where
$n_i\neq n_{i-1}$ (i.e. at radii where the azimuthal shift corresponding
to the third substep of the transport step is discontinuous). 
More generally
we want to examine the question of the continuity of $\xi^+_{ij}$
with respect to $\overline v_i^\theta\Delta t$. In order to check for this
continuity, we assume $\overline v_i^\theta = \left(N+\frac{1}{2}
+\epsilon\right)
\frac{\Delta y_i}{\Delta t}$, where $N$ is an integer, 
and we work out the behavior of 
$\xi_{ij}^+(\epsilon)$ in the vicinity of $\epsilon=0$. 
Since we have to use the explicit form of the
``$*/v^{\theta a}$'' operator, we adopt the van Leer algorithm (van
Leer, 1977), which
is widely used. Some straightforward algebra leads to~:

\[\xi_{ij}^+=\xi_{ij-N-1}^c+\left(\frac{1}{2}-\epsilon\right)
(\xi_{ij-N}^c-\xi_{ij-N-1}^c)\]

\begin{equation}
\label{eqn:conti}
\phantom{\xi_{ij}^+=}
-\left(\frac{1}{4}-\epsilon^2\right)
\frac{\Delta y_i}{2}(d\xi_{ij-N}^c-d\xi_{ij-N-1}^c)
\end{equation}
both for
$\epsilon>0$ and $\epsilon<0$ provided $|\epsilon| < \frac{1}{2}$ and where the operator ``$d\xi$'' is
the van Leer slope. Eq.~(\ref{eqn:conti})
shows that the field $\xi_{ij}^+$
is a continuous function of $\epsilon$ and hence of
$\overline v_i^\theta$. In particular no
special problem is to be expected from the discontinuities of $n_i$
across the disk.

\subsection{Operators swapping}

As we said in section~\ref{sec:clas}, it is a common practice to
alternate the radial $R$ and azimuthal $T$ transport 
operators every other timestep. In this modified algorithm, $R$
should usually be applied first, unless the velocity field is
updated just after applying the $T$ operator from
the new momenta and new density fields, or unless special
care is devoted to the $j$ indices. Indeed swapping blindly the $R$
and $T$ operators would result in moving radially the matter
with the radial velocity it actually has $\sim n_i$ cells upwards,
and would quickly end in a non-physical staggering everywhere
$n_i\neq 0$.

\section{Mono-dimensional tests}
\label{sec:1d}
\subsection{General considerations}
\label{referee2}

In order to validate this modified transport algorithm, we present
some 1D tests, and we compare the results of the standard method
and of the FARGO method on a realistic test problem. We solve simultaneously
the continuity and Navier Stokes equation for an isothermal gas (which
has a non-vanishing but small kinematic viscosity):
\new{
\begin{equation}
  \label{eq:conti1}
  \frac{\partial\rho}{\partial t}+\frac{\partial(\rho v)}{\partial x}=0
\end{equation}

\begin{equation}
  \label{eq:navsto1}
  \frac{\partial v}{\partial t}+v\frac{\partial v}{\partial x}=
-\frac{c_s^2}{\rho}\frac{\partial \rho}{\partial x}+\nu
\frac{\partial^2v}{\partial x^2}
\end{equation}

We assume that at rest the system has a uniform density $\rho_0$
and sound speed $c_s$.
The waves which can propagate in this system have the following
dispersion relationship:

\[  \omega=\pm\sqrt{k^2c_s^2-\frac{k^4\nu^4}{4}}-i\frac{k^2\nu}{2}\]

\begin{equation}
  \label{eq:rd1}
\mbox{~~~~or:~~~~}\omega=\pm kc_s-i\frac{k^2\nu}{2}\mbox{~~~if~}\nu\ll\nu_{\rm lim}=\frac{2c_s}{k}
\end{equation}

which reduces to the standard dispersion relation for an undamped 
acoustic  wave
$\omega=\pm kc_s$ provided the system is evolved for a time
small compared to the damping timescale $\tau = \frac{2}{\nu k^2}$.
This will be the case for the results we are going to present below,
so that any apparent damping of the waves has a numerical origin.
We do the following:
\begin{enumerate}
\item We first analyze the propagation of a sound wave in the matter
frame, \ie we take as initial conditions:
\begin{equation}
  \label{eq:ini1}
  \rho(x)=s\rho_0\cos(kx)\mbox{~~~and~~~}v(x)=sc_s\cos(kx)
\end{equation}
where $s$ is the wave relative amplitude. The polarization adopted
corresponds to a rightwards propagating wave. According to 
eq.~(\ref{eq:rd1}), it propagates with a phase velocity
which is $\Re \left(\frac{\omega}{k}\right)=c_s$. We study this
propagation with the standard transport algorithm (we are in the
matter frame so there is no
systematic average $x$-velocity, hence no need for a FARGO algorithm).
We check that in this case the solution we get is accurate
by varying the timestep and checking that the solution has converged.
\item We then turn to a case where the setup is slightly modified.
We take:
\begin{equation}
  \label{eq:ini1}
  \rho(x)=s\rho_0\cos(kx)\mbox{~~~and~~~}v(x)=v_0+sc_s\cos(kx)
\end{equation}
where $v_0$ is a constant, which we choose much bigger than $c_s$
(which would correspond to the conditions of a thin keplerian disk,
for example).
The evolution of the system from this setup ought to be the same as before,
since it merely corresponds to the same physical
situation, but described from a
frame moving at a constant speed $-v_0$ w.r.t the first one, so one
can invoke Galilean invariance to conclude that the wave profile 
evolution has to be the same. So any ``good'' algorithm should approach
as closely as possible the results of the matter frame simulations.
We show that this is not quite the case with the standard transport
method, which suffers from quite a high numerical dissipation, whereas
FARGO behaves much better (not to mention its much faster execution).
As a side result we also show that in this problem taking a CFL effective
ratio (for the standard transport method) bigger than $\frac{1}{2}$ leads
to an artificial and non-linear increase of the wave profile, and hence has
to be avoided.
\end{enumerate}

}

\subsection{1D Numerical results}
\begin{figure}[htbp]
  \psfig{file=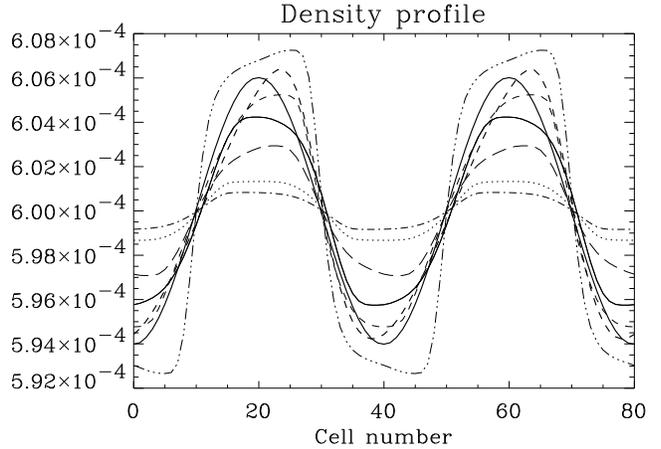,width=\columnwidth}
  \caption{Compared evolution of an acoustic wave evolved 
with the standard
transport algorithm and with the modified transport algorithm. We
plot only two of the five wavelengths, i.e. $80$ cells out of $200$.
\new{Due to numerical effects the phase velocity of all these profiles
do not {\em exactly} coincide with $c_s$, so that after a time $t_0$ their
phases do not coincide .
For this reason the profiles
have been shifted so that they have all approximately the same phase
in order to improve the clarity of the plot.}}
  \label{fig:res}
\end{figure}

We deal with a
1D grid composed of $N_s=200$ cells, with periodic boundary conditions.
The cell width is $\Delta x=0.0314$, 
the isothermal sound speed is $c_s=0.04$.
The equilibrium density is $\Sigma_0=6\cdot 10^{-4}$. These parameters
correspond roughly to the ones used in the numerical study of
a protoplanet on a circular orbit at 5~A.U.
embedded in a minimum mass protoplanetary disk
(Hayashi \etal 1985 or Bryden \etal 1998),
that are described in section~\ref{sec:2d}),
when the central star mass and the protoplanet orbit radius are taken
to be respectively
the units of mass and distance. We present 
the results of different test runs 
in fig.~\ref{fig:res}. The thick solid line represents
the initial profile, which corresponds to a rightward propagating
acoustic wave, with wavelength $\lambda=40\Delta x=1.256$. 
The relative amplitude of this
sound wave is $s=10^{-2}$. The thick dashed line represents
the density profile at time $t_0=220$, i.e. after the wave has
traveled $c_st_0/\lambda=7$ times its own wavelength, when studied
in the matter frame, i.e. when the velocity at $t=0$ is set to
be only the perturbed velocity associated to the sound wave. The
thick dashed profile is obtained with the \new{standard} transport
algorithm (there is no need for the modified one in this
case since we work in the matter frame), with a timestep 
$\Delta t=5\cdot 10^{-3}$. The curves obtained by choosing a
much smaller timestep appear to coincide exactly with this one, 
hence we can consider this thick dashed line as the actual state
the system must have at the date $t_0$. This profile does not
exactly coincide with the initial one because $t_0$ is 
$\sim\frac{1}{7}$ of the profile steepening time 
$t_{ps}\sim\frac{\lambda}{2c_ss}$.

Now if we just change
the initial velocity by uniformly adding $1.0$ to them
at $t=0$, which means that we are no more in the matter frame, and
we still work with the standard transport algorithm,
then we get the dotted profile, which has $\sim 1/5$ the amplitude
obtained from the computation in the matter frame. In this run the
CFL ratio is $v\Delta t/\Delta x= 0.16$. In order to check the 
timestep dependency of this result, we redo this test with twice
as smaller a timestep ($\Delta t= 2.5\cdot 10^{-3}$) and we get
the dash-dotted profile, which has about twice as smaller a density
contrast than the previous curve. Note that if this effect were
to be due to a physical kinematic viscosity $\nu$, then its value
should be~: $\nu\sim \frac{\lambda^2\log 5}{2\pi^2 t_0}\sim 5.8\cdot 10^{-4}$,
much higher than the expected viscosity in a minimum mass protoplanetary
disk ($\nu \sim 10^{-5}$ in our dimensionless units). 
Now, instead of decreasing the timestep, we increase it and set
$\Delta t=2.0\cdot 10^{-2}$ (hence the CFL ratio is about $0.64$).
We then get at time $t_0$ the dot-dot-dot-dashed profile, which is
not numerically damped but slightly amplified. With such a large
timestep, we can use the modified transport algorithm, which in
that case corresponds to a rightwards one cell shift and a leftwards
normal transport with a remaining CFL ratio of $1-0.64=0.36$. In that case
we get the thin long-dashed profile. If we use the modified FARGO transport
algorithm, we can still increase the timestep. The thin solid
profile and the thin short-dashed profile have been obtained 
respectively with $\Delta t=4\cdot 10^{-2}$ 
(effective CFL ratio $\sim 1.3$) 
and $\Delta t=1.2\cdot 10^{-1}$ (effective CFL ratio $\sim 3.8$).
We clearly see from these results that the FARGO transport
algorithm leads to less numerical dissipation than the standard
transport. 
From the
first two tests in the non-comoving frame, one can conclude that
increasing the \emph{number} of timesteps over a given
time interval with the standard transport algorithm
increases the numerical dissipation (if the
grid is moving w.r.t the matter frame with a
velocity $v_0\neq 0$ and if the main part of the velocity
comes from $v_0$). A simple explanation for the
lower numerical dissipation of the FARGO algorithm is
that
it requires less iterations as the timestep increases,
and since most of the  distance swept is achieved through an
exact shift (a circular permutation), the numerical dissipation
has to decrease as the timestep increases.

\section{Two-dimensional example~: the embedded protoplanet problem}
\label{sec:2d}

\begin{figure}[htbp]
  \begin{center}
    \psfig{file=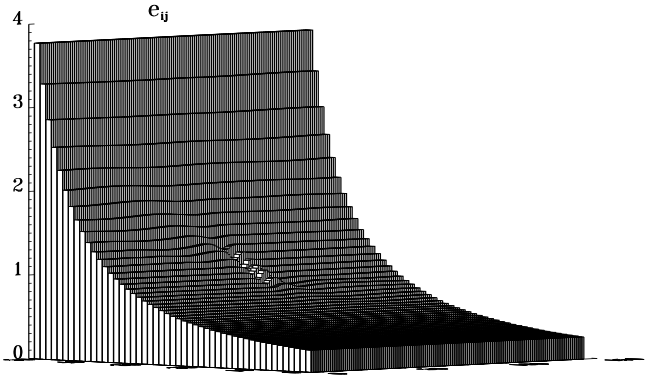,width=\columnwidth,height=5cm}
    \caption{Number of cells crossed during one timestep. See text for
parameters. The inner boundary is at the left (high values) and the outer boundary
at the right (low values).}
    \label{fig:eij}
  \end{center}
\end{figure}

We show in this section the validity  of the modified transport
algorithm when applied to the interaction of a Jupiter sized protoplanet
with a minimum mass protoplanetary disk in which it is
embedded. The perturbed potential
associated with the planet excites spiral density waves in the
disk, which propagate away both inwards and outwards, with a pattern
frequency equal to the planet orbital frequency.
The spiral waves interact with the disk and give it the angular
momentum they removed from the planet, and eventually open a gap
centered on the planet orbit, provided the planet mass is high
enough (Papaloizou \& Lin, 1984).
We present a run with a one solar
mass primary, one Jupiter mass protoplanet initially
on a fixed 
circular orbit at $r_0=5$~A.U. embedded in a standard protoplanetary
nebula whose parameters have been mentioned above. 
The grid has an inner
radius at $2$~A.U. and an outer radius at $12.5$~A.U. The sequence
$(R_i)_{i\in[0,N_r]}$ is equally spaced, with $N_r=49$~; The grid
has $N_s=143$ sectors, it is fixed in a non-Galilean
non-rotating frame centered
on the primary. Its
outer boundary is rigid and its inner boundary allows outflow but
no inflow. The disk aspect ratio is set to $4\cdot 10^{-2}$ everywhere.
The planet perturbed potential is smoothed on a length scale which amounts
to 40~\% of the Roche radius. In eq.~(\ref{eqn:shearlimit}) we
choose $C_0=0.5$.
We plot on fig.~\ref{fig:eij} the quantity $e_{ij}=\frac{v^{\theta
a}_{ij}\Delta t}{\Delta y_i}$ after $2.86$ orbits. This quantity
represents the effective CFL ratio.
With the standard transport algorithm this ratio
is bounded by $C_0$.

We see that the innermost ring sweeps almost four cells on one
timestep, hence the use of the FARGO transport algorithm in this
case results in a speed-up by a factor $\sim 8$  of the computation.
One can note that the difference in $e_{ij}$ between
the innermost ring and its immediate neighbor is $0.5$, which
is the maximum allowed by eq.~(\ref{eqn:shearlimit}). Indeed
the timestep in this run is shear-limited, and the constraint
on the residual velocities only
would lead to an even bigger timestep, since
as one can see the residuals of the distance swept over one
timestep amounts to far less than $1/2$, even in the vicinity
of the planet. 
\new{Indeed, runs performed with a logarithmic polar grid (\ie with
$R_{i+1}/R_i$ constant), which have a smaller value of $R_{i+1}-R_i$
in the inner part, have shown to allow a speed-up by a factor $\sim 30$
w.r.t. the standard method.}

\begin{figure}[htbp]
  \begin{center}
    \psfig{file=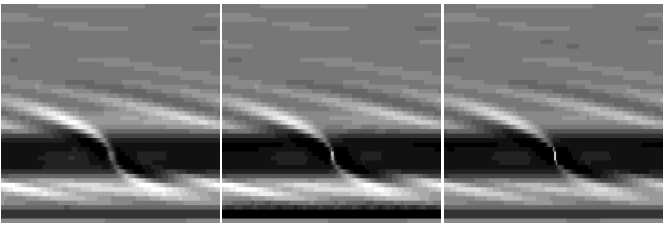,width=\columnwidth}
    \caption{Disk density $\Sigma_{ij}$; $j$ is in abscissa and
$i$ in ordinate. The left plot has been obtained by a \new{non-rotating} frame
standard method, the middle one by a \new{non-rotating} 
frame FARGO transport method
and the right one by a corotating frame standard method. 
\new{Since each of these
plots is approximately square, any circular feature in the disk should
appear on the plots as a $1:3$ vertical ellipse. This is not quite
the case of the material surrounding the planet 
in the left panel, which leads to the conclusion that in a
non-rotating frame standard transport method, the matter is artificially
elongated along the orbital motion. The FARGO case, in the middle panel,
shows much better behavior, and the coorbital material
has a distribution which looks very much like the right panel one.}
}
    \label{fig:diskdens}    
  \end{center}
\end{figure}

\begin{figure}[htbp]
  \begin{center}
    \psfig{file=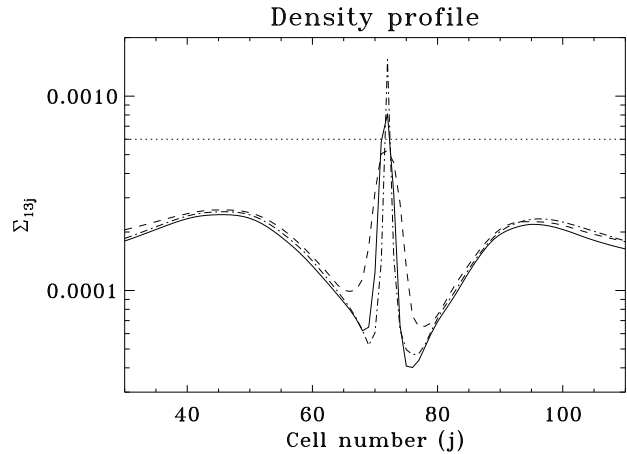,width=\columnwidth}
    \caption{Disk density cuts at the planet radius.
 The solid line represents
      the FARGO transport case, the dashed line represents 
the standard case,
and the dot-dashed line represents the corotating frame result.
The dotted line indicates the unperturbed surface density. Note that
the local maxima at $j\simeq 46$ and $j\simeq  94$ correspond to 
a temporary
residual accumulation of material at the $L_4$ and $L_5$ Lagrange
points of the protoplanet. }
    \label{fig:denspeak}
  \end{center}
\end{figure}

In order to see how numerical viscosity affects the disk response
in both cases, we plot on the fig.~\ref{fig:diskdens} the disk density
after $28.6$ orbits, obtained from different algorithms. The left plot
corresponds to a non-rotating frame standard transport run, while the middle
plot represents a non-rotating frame FARGO transport run, and the
right plot represents a standard transport run in a frame corotating
with the planet (hence the planet is fixed with respect to the grid,
so we expect from the results of section~\ref{sec:1d} the density response
in the vicinity of the planet to be given with a high accuracy). Note
that  special care has to be devoted to the treatment of the Coriolis
force in that case in order to conserve exactly the angular momentum
and then to avoid a spurious outwards transport in the disk (Kley, 1998).
We clearly see that the global spiral pattern excited by the protoplanet
in the disk is identical in the three cases, though the response in
the immediate vicinity of the planet is much more spread out in the
\new{non-rotating} frame standard accretion case (left plot), and that the most
sharply peaked response is achieved through the use of a corotating frame
(right plot), as expected.
Indeed, we plot on fig.~\ref{fig:denspeak} a
cut of the disk density at the planet radius in the three cases.
The solid line represents the FARGO transport result, and the
dot-dashed line the corotating frame result. They both have the same
width, though the maximum of the density in the corotating case is
higher. The dashed curve represents the result of the standard
transport in a \new{non-rotating} frame. Its width is about twice as large as the
other curves' width, and we also see that numerical effects in that case
lead to additional 
leading and trailing material (near cells number~$65$ and $77$),
and to a smaller density peak value.

\begin{figure}[htbp]
  \begin{center}
    \psfig{file=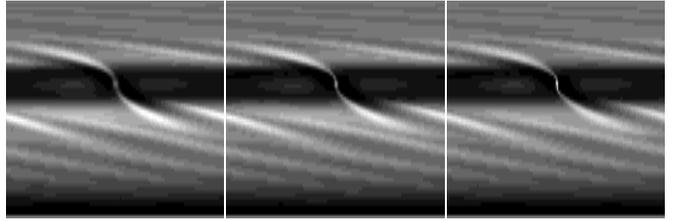,width=\columnwidth}
    \caption{\new{Disk density $\Sigma_{ij}$; $j$ is in abscissa and
$i$ in ordinate, for the log-grid runs described in the text.
 The left plot has been obtained by a \new{non-rotating} frame
standard method, the middle one by a \new{non-rotating} 
frame FARGO transport method
and the right one by a corotating frame standard method. 
The same comments as in figure~\ref{fig:diskdens} apply here. On
this specific example, the FARGO run turned out to be $17$~times faster
than the standard run in the non-rotating frame, and $15$~times faster
than the standard run in the co-rotating frame.}
}
    \label{fig:loggrid}    
  \end{center}
\end{figure}

\new{The FARGO plot on figure~\ref{fig:diskdens}
exhibits at its inner boundary an
oscillatory behavior  which originates from three combined effects.
First, this is a shear-limited run ---~see eq.~(\ref{eqn:shearlimit})
and fig.~\ref{fig:eij}~---; if we change the $0.5$ factor in 
eq.~(\ref{eqn:shearlimit}) to $0.3$, this oscillatory behavior disappears
(hence in any high resolution run, where the algorithm is most likely
to be residual velocity limited rather than shear-limited, it never
turns up).
Second, the inner grid has strongly radially elongated cells. If we take
a log-grid (see e.g. Nelson, 1999, or the example below), 
where the cells are almost ``square''
everywhere, this behavior is not observed, even if the run remains
shear-limited. And finally, 
we have a
steep density gradient close to the inner boundary. If the inner boundary
was closed and hence if we had no density gradient, this oscillatory
behavior would never appear. In all the cases where it was observed
this behavior always
disappeared after a few tens of dynamical times.

It should
be noted that the numerical damping observed in the non-stationery
frame in section~\ref{sec:1d} occurs both in the non-rotating frame
{\em and corotating frame (far from the coorbital region)} 
standard method runs. Hence the amplitude of the
protoplanet triggered density wave is marginally higher in a FARGO
run at the
inner boundary. Both this reason and the effect we noticed in the
previous paragraph
 lead to a marginally higher mass loss through
the inner boundary, at least during the first stages of the evolution
of the system, which results in the darker band at the inner boundary
 in the middle panel of
figure~\ref{fig:diskdens}.

We present in figure~\ref{fig:loggrid} the results of three runs
(non-rotating standard and FARGO, and corotating standard), which
describe the same physical system as before after the same amount
of time, but with a grid for which $N_r=70$, $N_s=180$, $R_{min}=0.25$
and $R_{max}=2.5$, and with a geometric sequence for 
$(R_i)_{i\in [0,N_r]}$ (hence it is a log-grid, and everywhere its cells are
almost ``square''). One can check on these plots that there
is no oscillatory behavior in the FARGO results (this time the cells are no 
radially elongated near the inner boundary), whereas the
run is still shear-limited. Furthermore, as stated
above, a careful look at the  inner spiral structure shows that it
has a slightly higher amplitude in
the FARGO case.
}

\new{From the results depicted in 
figures~\ref{fig:diskdens}, \ref{fig:loggrid}
and~\ref{fig:denspeak},} one can deduce that the FARGO transport algorithm
on this particular problem is much closer than
the usual standard transport algorithm to the exact solution
(which must closely resemble the results given by the corotating
frame run, at least in the coorbital region, since in section~\ref{sec:1d}
we have seen that one needs to be in the comoving frame in order to
get accurate results even in the limit of a vanishing timestep). 
{
Another quantitative evaluation of the FARGO algorithm consists in monitoring
the accretion rate onto the planet as a function of time. We present on
figure~\ref{fig:monitoracc} the accretion rate onto a one Jupiter mass 
protoplanet embedded in a minimum mass protostellar disk with 
no initial gap. The
disk parameters are the same as before, as well as the grid resolution
(\new{arithmetic radial spacing with $N_r=49$ and $N_s=143$}). 
Three runs are presented with
three different schemes: the \new{standard}
method in the rotating frame, which
gives, according to section~\ref{sec:1d}, the
most accurate results, the \new{standard} method in the \new{non-rotating} frame, and the
\new{FARGO} method in 
the \new{non-rotating} frame. \new{We use a slightly different accretion
procedure than the one 
described by Kley (1999)}. 
\begin{figure}[htbp]
  \begin{center}
    \psfig{file=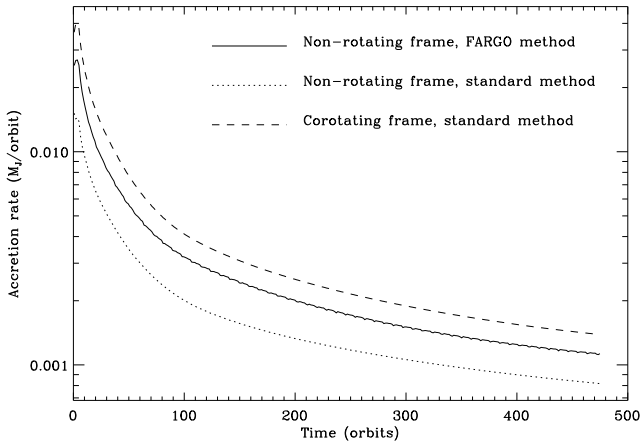,width=\columnwidth}
    \caption{Accretion rate as a function of time onto a one
Jupiter mass protoplanet with three different methods. See text
for details.}
    \label{fig:monitoracc}
  \end{center}
\end{figure}
We see from the curve obtained in the
\new{corotating} frame that the accretion rate is about 
$1.6\cdot 10^{-3}$~$M_J.\mbox{orbit}^{-1}$ after $400$ orbits. This
is in relatively good agreement with Kley's results, who gets slightly
more than $2.0\cdot 10^{-3}$~$M_J.\mbox{orbit}^{-1}$ after $400$ orbits
in a similar run, but with a different grid resolution \new{and a
slightly different accretion protocol}.
We see from figure~\ref{fig:monitoracc} that
the accretion rate
in the \new{non-rotating} 
frame, with a \new{standard} method, is smaller than in the rotating
frame run, by a factor $\simeq 2$. The fact that the accretion
rate is slower in this case was to be expected from the
curves of fig.~\ref{fig:denspeak}. Now the run with the \new{FARGO} 
algorithm
leads to an accretion rate which is between the rotating frame results
and the \new{non-rotating} frame \new{standard} 
transport results, and which are closer to
the rotating frame results. From these considerations again we see that
the \new{FARGO} transport leads to a smaller error w.r.t. the rotating frame
results.
}
The point here is that the
FARGO transport algorithm is about one order of magnitude or more 
faster than
the corotating frame standard transport run, and that the corotating frame
is suitable only to the study of a protoplanet on a fixed circular
orbit. From these remarks it clearly appears that the FARGO transport
algorithm is particularly well suited to the study of the protoplanet
orbit long-term evolution. \new{FARGO has already been used to study
the migration and mass accretion of a Jupiter sized protoplanet in
a protoplanetary disk. It has been extensively tested against existing
independent codes, which use the standard transport algorithm. It has
proven to give very similar results, and the slight differences which
remain between these codes and FARGO can all be understood in terms
of FARGO's lower numerical diffusivity (Nelson \etal 1999).}

\section{Conclusion}

The FARGO algorithm for the azimuthal transport turns out to 
be able to speed up by about an order of magnitude the numerical
simulation of a differentially rotating disk, with a smaller numerical
viscosity than the usual transport algorithm. It has been validated
by many tests on the embedded protoplanet problem.
It is worth mentioning that
the FARGO transport algorithm must be used with a good understanding
of the physical processes at work in the system. In particular,
the timestep given by eq.~(\ref{eqn:timestep}) must be short
compared to all the physical time scales relevant for the system.
\new{In the case we have presented in this paper this is automatically
ensured by the set of eq.~(\ref{eqn:timelimit}) to~(\ref{eqn:timestep}),
but if additional physics is to be added (magnetic field, radiative
transfer, etc.), the timestep limit needs to be carefully worked out.}
Furthermore, \new{no advantage is gained in using} FARGO 
in problems where the perturbed
velocity is comparable to the rest velocity. It is the case
for instance of the gas flow in a galactic bar. 
\new{This does not mean that the FARGO algorithm leads to
wrong results in that case, but simply that it will not be better than a 
standard method, both in terms of numerical diffusivity and
execution time.}
On the other
hand, the FARGO algorithm appears to be very
well suited to all the cases where
the perturbed velocities in any differentially rotating disk are small
compared to the unperturbed velocities, \new{which does not mean that
the problem under consideration has to be linear; indeed the relative
perturbed amplitude can be arbitrarily high (see \eg section~\ref{sec:2d}
in which the protoplanet wake generates shocks in the disk).}
More generally the FARGO
algorithm can be used to describe the HD evolution of any sheared fluid
on a fixed orthogonal eulerian grid.

\section{Acknowledgments}

I am indebted to Richard P. Nelson for many valuable
discussions and  suggestions in the course of this
work, as well as for a careful reading of the successive drafts of
this paper, to Jim M. Stone for advice on eulerian numerical simulations,
to John C. B. Papaloizou for many  discussions
on the protoplanet migration problem, to Willy Kley for advice
on the corotating frame runs,
\new{and to an anonymous referee whose comments led to 
an improvement of the paper.}
This work was done in the research
network ``Accretion onto black holes, compact stars and protostars''
funded by the European Commission under contract number ERBFMRX-CT98-0195.


\end{document}